\documentclass{article}

\usepackage[preprint]{neurips_2026}

\usepackage[utf8]{inputenc}
\usepackage[T1]{fontenc}
\usepackage{graphicx}
\usepackage{amsmath}
\usepackage{amssymb}
\usepackage{amsfonts}
\usepackage{booktabs}
\usepackage{multirow}
\usepackage{nicefrac}
\usepackage{microtype}
\usepackage{xcolor}
\usepackage{listings}
\usepackage{subcaption}
\usepackage{float}
\usepackage{enumitem}
\usepackage{hyperref}
\usepackage{url}

\title{Evidence Units:\\
Ontology-Grounded Document Organization\\
for Parser-Independent Retrieval}

\author{%
  Yeonjee Han \\
  GenApp Tech, KT (Korea Telecom) \\
  \texttt{yeonjee.han@kt.com} \\
  \url{https://github.com/hanyeonjee/evidence-units}
}

\begin{document}

\maketitle

\begin{abstract}
Structured documents---tables paired with captions, figures with explanations,
equations with the paragraphs that interpret them---are routinely fragmented when
indexed for retrieval. The dominant element-level indexing paradigm treats every
parsed element as an independent chunk, scattering semantically cohesive units
across separate retrieval candidates. This fragmentation forces systems to find
multiple disjoint pieces to answer a single question, even when all the evidence
appears on the same page. This paper presents a parser-independent pipeline that
constructs \textbf{Evidence Units (EUs)}: semantically complete document chunks that
group visual assets with their contextual text.

Our approach introduces four contributions: (1)~an \textbf{ontology-grounded canonical
role normalization} extending DoCO (Document Components Ontology) that maps heterogeneous
parser outputs to a unified semantic schema; (2)~a \textbf{semantic global assignment
algorithm} that optimally assigns paragraphs to EUs via a full similarity matrix; (3)~a
\textbf{graph-based decision layer} stored in Neo4j that formalizes EU construction rules
and validates EU completeness through two invariants; and (4)~\textbf{cross-parser
validation} demonstrating that EU spatial footprints converge across structurally
different parsers (MinerU, Docling), with EU performance gains preserved even under
parser-induced bbox variance.

Experiments on a filtered subset of OmniDocBench v1.0 (1,340 pages retained from 1,355
after removing non-content-only pages; 1,551 QA pairs) with ground-truth annotations
show EU-based chunking improves retrieval LCS by \textbf{+0.31} over element-level
chunking ($0.50\!\to\!0.81$). Recall@1 increases from 0.15 to \textbf{0.51}
($3.4\times$), and MinK decreases from 2.58 to \textbf{1.72}. Cross-parser experiments
confirm the relative gain ($\Delta$LCS\,$\approx$\,+0.23--+0.31) is preserved across
MinerU and Docling, validating parser independence empirically. Text queries show the
most dramatic gain: Recall@1 rises from 0.08 to \textbf{0.47}.
\end{abstract}

\section{Introduction}
\label{sec:intro}

RAG systems chunk documents into retrievable units. The chunking strategy directly
impacts whether the retrieved context contains the information needed to answer a query.
Current methods have three fundamental limitations:

\textbf{Semantic Fragmentation.} Fixed-size chunking (e.g., 1024 tokens) and
element-level chunking separate semantically coupled elements. A table's HTML markup,
its caption ``Table~3: Revenue by Segment,'' the unit label ``(Unit: million USD),'' and
the explanatory paragraph become independent chunks. When a user asks ``What was the
total revenue?'', the retriever finds the caption (high text similarity) but not the
table data (low similarity with the query).

\textbf{Parser Dependence.} The same document produces not only different labels
but also structurally different bounding box (bbox) representations across parsers.
For the same table region, Parser A emits a single bbox, Docling splits it into three
row-level bboxes, and PaddleOCR-VL (a Vision-Language Model) produces a single bbox
with inference-induced positional error of $\pm$0.02 in normalized coordinates.
Without normalization, chunking algorithms must be rewritten for each parser---and even
the spatial grouping logic must account for each parser's unique bbox decomposition.

\textbf{No Quality Assurance.} Generated chunks are not validated for semantic
completeness. A table chunk without any contextual anchor is semantically incomplete but
passes through unchecked.

We propose \textbf{Evidence Units (EUs)}: parser-independent, semantically complete
document chunks constructed through a three-stage pipeline:
\begin{enumerate}[leftmargin=*]
  \item \textbf{Ontology-Grounded Node Normalization}: Parser outputs are mapped to
        canonical roles defined in a DoCO-extended OWL ontology, decoupling EU
        construction from parser specifics.
  \item \textbf{EU Construction (3 Phases)}: Visual seeds anchor EU formation;
        structural elements attach by proximity (Phase~A); paragraphs attach by global
        semantic similarity (Phase~B); residual elements consolidate by layout-driven
        heuristics (Phase~C).
  \item \textbf{Graph-Based Decision Layer}: D1 Construction rules are realized inline
        in the EU generation pipeline, with formal definitions persisted in Neo4j as the
        authoritative specification. D2/D3 invariants are encoded as graph-resident
        schemas for future runtime enforcement.
\end{enumerate}

A key property enabling parser independence is \textbf{EU spatial footprint
convergence}: although individual element bboxes differ across parsers, the bounding
box of the complete EU (the union of all its member bboxes) converges to the same
document region. We formalize and empirically verify this property in
Section~\ref{sec:footprint}.

\paragraph{Contributions.}
\begin{enumerate}[leftmargin=*]
  \item \textbf{Ontology-grounded canonical role system} (DSO, extending DoCO/SPAR)
        with \texttt{skos:altLabel} aliases enabling parser-independent normalization and
        embedding fallback for unknown parsers
  \item \textbf{Global semantic allocation (Phase~B)} computing a full
        paragraph$\times$EU similarity matrix for optimal assignment---fundamentally
        different from prior local 1:1 proximity matching
  \item \textbf{Graph-based decision layer} storing D1 construction rules as Neo4j nodes
        with NEXT chains, with D2/D3 invariant schemas persisted for future runtime
        enforcement
  \item \textbf{EU spatial footprint convergence} formalized and empirically verified
        across MinerU and Docling on OmniDocBench (1,340 usable pages), showing consistent
        $\Delta$LCS\,$\approx$\,+0.27 regardless of parser choice
\end{enumerate}

\section{Related Work}
\label{sec:related}

\subsection{Document Chunking for RAG}

\textbf{Fixed-size splitting} (LlamaIndex SimpleNodeParser, LangChain
RecursiveCharacterTextSplitter) divides text by token count, ignoring document
structure. \textbf{Semantic chunking}~\cite{kamradt} splits at embedding similarity
discontinuities but does not model visual elements. \textbf{Layout-aware chunking}
(Unstructured.io) respects element boundaries but treats each element independently.

\textbf{S2 Chunking}~\cite{s2chunk} constructs a weighted graph using bounding box
coordinates and text embeddings, then applies spectral clustering. \textbf{SCAN}~\cite{scan}
groups document components into semantically coherent regions for VLM-friendly RAG.
\textbf{Vision-Guided Chunking}~\cite{vgchunk} uses multimodal understanding to enhance
chunk boundaries. These methods share a local matching pattern: elements are linked when
both distance and semantic conditions are met (e.g., distance threshold \emph{AND}
embedding similarity $\geq 0.7$). Critically, all assume a single, fixed parser; none
address the structural bbox variation that arises when the same parser is replaced.

\textbf{Docling}~\cite{docling} (IBM, 2024) is a document parsing library that provides
two RAG-oriented chunkers: \texttt{HierarchicalChunker}, which follows the inferred
section tree, and \texttt{HybridChunker}, which combines hierarchy with token-length
limits. Both integrate with LlamaIndex and LangChain. However, Docling chunking is
\emph{parser-specific}---it operates only on Docling's own intermediate representation
and produces no output when Docling fails to parse a page. More fundamentally, it does
not guarantee the semantic co-location of visual elements: a table and its caption may
appear in different chunks depending on how the section hierarchy is reconstructed.
EU construction, by contrast, operates on top of \emph{any} parser output via canonical
role normalization, and explicitly enforces visual-semantic grouping through a typed
ontology (\texttt{table\_panel} $=$ table $+$ caption $+$ unit\_label). Table~\ref{tab:related} summarizes these distinctions. Our experiments in Section 5.3 include Docling output as one parser track, directly measuring the gain of EU construction on top of Docling's own chunking.

\begin{table}[h]
\centering
\caption{Comparison with prior chunking methods.}
\label{tab:related}
\small
\begin{tabular}{lll}
\toprule
\textbf{Aspect} & \textbf{Prior Methods~\cite{s2chunk,scan,vgchunk,docling}} & \textbf{This Work (EU)} \\
\midrule
Parser input         & Single parser, fixed            & Multi-parser auto-normalization \\
Bbox representation  & Assumed consistent              & Converged via EU footprint \\
Paragraph assignment & Local neighborhood matching     & \textbf{Global similarity matrix} \\
Caption+table pairing & Implicit / not guaranteed      & \textbf{Explicit ontology (typed EU)} \\
Unmatched handling   & Often discarded                 & Preserved as independent EU \\
Quality validation   & None or minimal                 & \textbf{2-Invariant} (I1, I2) \\
Rule management      & Hardcoded thresholds            & \textbf{Neo4j dynamic rule chain} \\
Output               & Flat chunks                     & Structured EU with metadata \\
\bottomrule
\end{tabular}
\end{table}

\subsection{Graph-Based RAG}

\textbf{GraphRAG}~\cite{graphrag} (Microsoft, 2024) builds knowledge graphs from document
\emph{content} (entities, relations) to enhance retrieval via community summaries.
\textbf{KG-RAG}~\cite{kgrag} integrates biomedical knowledge graphs with RAG.
These approaches build graphs from content; we build a graph of document
\emph{structure} (layout elements, spatial relations, EU membership). The two are
complementary.

\subsection{RAG Evaluation}

\textbf{OHR-Bench}~\cite{ohrbench} (ICCV 2025) evaluates how OCR quality impacts RAG,
keeping chunking fixed at 1024 tokens. Our work is orthogonal: we hold parser quality
constant and vary the chunking strategy, addressing a complementary axis of RAG quality.
The cross-parser experiment in Section~\ref{sec:crossparser} additionally varies the
parser while keeping the chunking strategy fixed, isolating the contribution of EU
construction from parser accuracy effects.

\section{Method}
\label{sec:method}

\subsection{System Overview}

\begin{figure}[t]
\centering
\includegraphics[width=\linewidth]{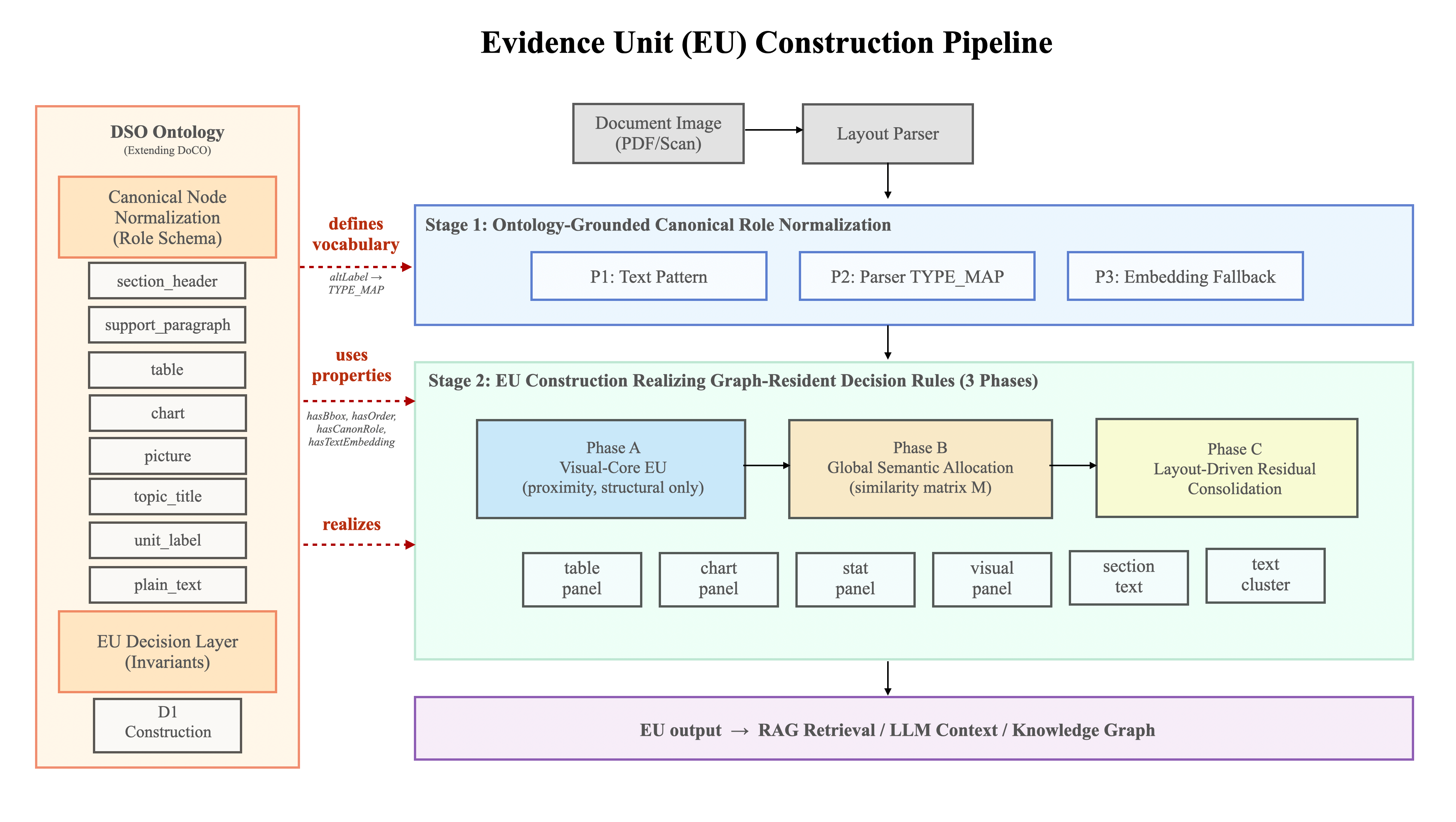}
\caption{EU Construction Pipeline. Stage~1: ontology-grounded node normalization via
cascaded role assignment (pattern matching $\to$ TYPE\_MAP $\to$ embedding fallback).
Stage~2: three-phase EU construction (Phase~A: visual-core formation; Phase~B: global
semantic allocation; Phase~C: residual consolidation). EU output feeds downstream
applications: RAG retrieval, LLM context assembly, and knowledge graph ingestion.}
\label{fig:pipeline}
\end{figure}

The pipeline has three stages: (1)~ontology-grounded canonical role normalization,
(2)~three-phase EU construction, and (3)~graph-based decision layer validation.
Figure~\ref{fig:pipeline} illustrates the full pipeline.

\subsection{Stage 1: Ontology-Grounded Node Normalization}
\label{sec:normalization}

\subsubsection{Document Structure Ontology (DSO, extending DoCO)}

Our ontology extends \textbf{DoCO} (Document Components Ontology)~\cite{doco}, an
OWL~2~DL ontology from the SPAR suite~\cite{spar}. We introduce
\texttt{dso:LayoutElement} as base class---distinct from \texttt{doco:Section} (``headed
container'')---representing any detected layout region. Each concrete class inherits from
both \texttt{dso:LayoutElement} (EU-specific properties) and the corresponding DoCO
class (semantic interoperability):

\noindent\begin{minipage}{\linewidth}
\begin{lstlisting}[language=]
@prefix doco: <http://purl.org/spar/doco/> .
@prefix dso:  <http://document-eu.org/ontology/dso#> .

dso:Paragraph     rdfs:subClassOf dso:LayoutElement, doco:Paragraph .
dso:Table         rdfs:subClassOf dso:LayoutElement, doco:Table .
dso:Chart         rdfs:subClassOf dso:LayoutElement, doco:FigureBox .
dso:Picture       rdfs:subClassOf dso:LayoutElement, doco:Figure .
dso:Caption       rdfs:subClassOf dso:LayoutElement, deo:Caption .
dso:SectionHeader rdfs:subClassOf    dso:LayoutElement ;
                  owl:equivalentClass doco:SectionTitle .
dso:SectionHeader skos:altLabel "SectionTitle"@en, "heading"@en,
                                 "section_header"@en, "header"@en .

% EU-specific properties (not in DoCO)
dso:hasBbox          a owl:DatatypeProperty .  % normalized [x1,y1,x2,y2] in [0,1]
dso:hasOrder         a owl:DatatypeProperty .  % reading order
dso:hasCanonRole     a owl:DatatypeProperty .  % canonical role
dso:hasTextEmbedding a owl:DatatypeProperty .  % 384-dim vector

% EvidenceUnit as first-class ontology concept
dso:EvidenceUnit a owl:Class .
dso:hasMember    a owl:ObjectProperty .  % EU -> LayoutElement
dso:euKind       a owl:DatatypeProperty .
\end{lstlisting}
\end{minipage}

All bboxes are stored in normalized $[0,1]$ coordinates (dividing pixel coordinates by
page width/height), making spatial distance thresholds parser- and resolution-agnostic.
\texttt{dso:SectionHeader} is declared \texttt{owl:equivalentClass} to
\texttt{doco:SectionTitle}, correctly positioning it above Paragraph but below
document-level Title. The \texttt{skos:altLabel} annotations enumerate known parser label
variants, providing \emph{ontology-level grounding} for the runtime TYPE\_MAP.

The EU generation pipeline \textbf{realizes} the D1 Construction rules by accessing DSO
properties directly: \texttt{hasBbox} and \texttt{hasOrder} drive spatial distance
(D1\_010) and header boundary gating (D1\_031); \texttt{hasCanonRole} governs
role-based branching; \texttt{hasTextEmbedding} drives Phase~B semantic assignment
($\tau=0.40$, D1\_040). EU-level constraints are enforced via \texttt{hasMember}
(D1\_021) and \texttt{euKind} (D1\_051).

\subsubsection{Three-Priority Normalization}

Canonical role assignment follows a strict priority order grounded in ontology
\texttt{altLabel} vocabulary:

\textbf{Priority~1---Text Pattern Matching} (parser-independent):
\begin{lstlisting}
"(Unit: million)" -> unit_label   % regex: ^\(\s*[Uu]nit\s*:
"[R&D Projects]"  -> topic_title  % regex: ^\[.+\]$
\end{lstlisting}

\textbf{Priority~2---Parser TYPE\_MAP} (derived from ontology \texttt{altLabel}s):
\begin{lstlisting}
Parser A   "SectionHeader" -> section_header  % altLabel "section_header"
Docling   "heading"       -> section_header  % altLabel "heading"
MinerU    "title"         -> section_header  % altLabel "title"
PaddleOCR "title"         -> section_header  % via embedding fallback
\end{lstlisting}

\textbf{Priority~3---Embedding Similarity Fallback} (unknown parsers):
\[
\text{cosine\_sim}(\text{raw\_label},\; \text{canon\_role}) \;\geq\; 0.80
\quad\Rightarrow\quad \text{assign canon\_role}
\]

\subsection{Stage 2: EU Construction (3 Phases)}
\label{sec:construction}

\subsubsection{Phase A: Visual-Core EU Formation}

Visual elements (table, chart, picture) become EU seeds. Only \emph{structural roles}
(section\_header, unit\_label, topic\_title) attach via spatial proximity:
\begin{equation}
d_{\text{spatial}}(v,\, n) \;=\; v_{\text{gap}}(v, n) + 0.3 \cdot x_{\text{diff}}(v, n) \;<\; 0.30
\end{equation}
\texttt{support\_paragraph} is explicitly excluded and handled semantically in
Phase~B. Adjacent visual EUs (Y-gap $<0.22$) within the same section merge into
\texttt{stat\_panel} (table + chart showing same data).

\subsubsection{Phase B: Global Semantic Allocation }

With visual-core EUs formed in Phase~A, Phase~B assigns \emph{contextual elements}
(\texttt{support\_paragraph}) via global semantic similarity rather than spatial
position. All unassigned paragraphs and all existing EUs form a similarity matrix
$\mathbf{M}$:
\begin{equation}
M_{ij} \;=\; \max_{k\,\in\,\mathrm{EU}_j}\;
\cos\!\left(\mathbf{e}_{\mathrm{para}_i},\;\mathbf{e}_{\mathrm{member}_k}\right)
\end{equation}
\begin{equation}
\mathrm{para}_i \;\longrightarrow\; \arg\max_j\, M_{ij}
\quad \text{if } \max_j M_{ij} \geq \tau \;(\tau = 0.40)
\end{equation}
Paragraphs with $\max_j M_{ij} < \tau$ are preserved as independent EUs (no
information loss). Table~\ref{tab:phaseb} contrasts this with prior local matching.

\begin{table}[h]
\centering
\caption{Phase~B global allocation vs.\ prior local matching.}
\label{tab:phaseb}
\small
\begin{tabular}{lll}
\toprule
\textbf{Property} & \textbf{Prior (Local 1:1)} & \textbf{Phase~B (Global)} \\
\midrule
Comparison scope & Pivot's 4 neighbors       & All paragraphs $\times$ all EUs \\
Assignment       & First match wins           & Optimal (argmax) \\
Threshold        & 0.70 (high, local)         & 0.40 (lower, global context) \\
Unmatched        & Discarded                  & Preserved as independent EU \\
\bottomrule
\end{tabular}
\end{table}

\subsubsection{Phase C: Layout-Driven Residual Consolidation}

Nodes unassigned after Phases A and B undergo three fallback operations:
\begin{itemize}[leftmargin=*]
  \item \textbf{C-1: Section-Delimited Grouping.} Each section\_header collects
        subsequent paragraphs until the next boundary.
  \item \textbf{C-2: Proximity-Based Label Reattachment.} Orphaned structural markers
        anchor to the spatially nearest visual EU ($\text{gravity\_dist} < 0.25$).
  \item \textbf{C-3: Residual Paragraph Clustering.} Remaining paragraphs consolidate
        by spatial contiguity (vertical gap $<0.07$, X-alignment $<0.18$, order gap
        $\leq 3$).
\end{itemize}

\subsection{Stage 3: Graph-Based Decision Layer}
\label{sec:decision}

\subsubsection{D1 Construction Rules as Graph-Resident Specification}

The D1 Construction rules are implemented inline in the EU generation pipeline, with
formal definitions persisted in Neo4j as the authoritative specification
(Table~\ref{tab:d1}):

\begin{table}[h]
\centering
\caption{D1 Construction rules realized in the EU generation pipeline.}
\label{tab:d1}
\small
\begin{tabular}{lll}
\toprule
\textbf{Rule} & \textbf{Name} & \textbf{Parameter} \\
\midrule
D1\_010 & Proximity-Based Structural Mapping  & \texttt{max\_gravity\_reach=0.30} \\
D1\_021 & Homogeneous Visual Exclusion        & table+chart allowed \\
D1\_031 & Section Boundary Gating            & order gap threshold \\
D1\_040 & Semantic Paragraph Attachment      & $\tau=0.40$ \\
D1\_051 & Type-Conflict Merge Guard          & same-type visual $\leq 1$ \\
\bottomrule
\end{tabular}
\end{table}

D2 Restoration and D3 Validation invariants are encoded as graph-resident schemas,
enabling future runtime enforcement without code modification.

\subsubsection{Rules as Knowledge Graph Nodes}

\noindent\begin{minipage}{\linewidth}
\begin{lstlisting}[language=]
(:DecisionLayer {name:'EU_Decision_Layer', version:'2.0'})
  -[:HAS_RULE]->
(:DecisionRule {rule_id:'D1_010', phase:'D1_CONSTRUCTION', active:true})
  -[:NEXT]->
(:DecisionRule {rule_id:'D1_040', phase:'D1_CONSTRUCTION', active:true})
  -[:NEXT]->
(:DecisionRule {rule_id:'D2_010', phase:'D2_RESTORATION', active:true})
  -[:NEXT]->
(:DecisionRule {rule_id:'D2_020', phase:'D2_RESTORATION', active:true})
\end{lstlisting}
\end{minipage}

\subsubsection{Invariant I1: Anchoring}
A visual EU must contain $\geq$1 anchor element (section\_header, unit\_label,
topic\_title). Violation: search for nearby unassigned anchor elements ($\text{gravity\_dist} < 0.30$); if no such anchor is found, the element is demoted to \texttt{plain\_text}.

\subsubsection{Invariant I2: Type Consistency}
A stat\_panel (merged table~+~chart) must represent the same data:
\[
\frac{|\text{table\_values} \cap \text{chart\_values}|}{|\text{chart\_values}|}
\;\geq\; 0.60
\]
Violation: split stat\_panel into separate table\_panel + chart\_panel.

\subsection{Parser Independence via EU Spatial Footprint Convergence}
\label{sec:footprint}

\subsubsection{Why Parser Outputs Differ Structurally}

\begin{figure}[t]
\centering
\includegraphics[width=\linewidth]{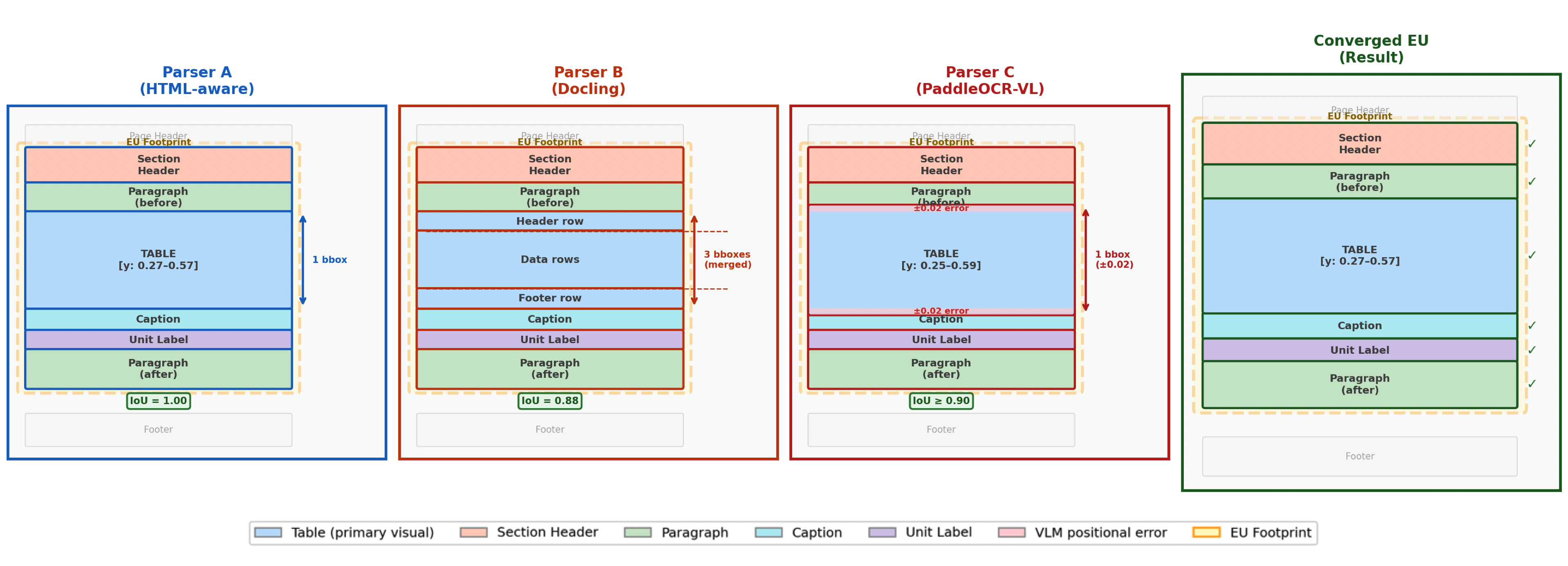}
\caption{EU spatial footprint convergence across parsers. Each column shows how a
different parser decomposes the same table region: Parser A emits one bbox; Docling
emits three row-level bboxes that merge in Phase~A; PaddleOCR-VL emits a single
VLM-inferred bbox with $\pm$0.02 positional error (shown in red). Despite these
differences, the EU footprint (yellow dashed outline)---the bounding box of all EU
members including section header, caption, unit label, and adjacent paragraphs---
converges to the same page region across all parsers. The Docling case achieves
IoU\,=\,0.88 due to an attachment range effect on the trailing paragraph; all other
parsers achieve IoU\,=\,1.00.}
\label{fig:footprint}
\end{figure}

Different parsers produce not merely different label strings but fundamentally different
\emph{spatial decompositions} of the same document region.
Figure~\ref{fig:footprint} illustrates this for a single table region:

\begin{itemize}[leftmargin=*]
  \item \textbf{Parser A} (HTML-aware): recognizes the full table with
        colspan/rowspan information and emits \emph{one} bbox covering the entire region
        $[y{:}0.27{\sim}0.57]$.
  \item \textbf{Docling} (layout detection model): splits the table into three
        row-level bboxes---header row $[y{:}0.27{\sim}0.33]$, data rows
        $[y{:}0.33{\sim}0.51]$, footer row $[y{:}0.51{\sim}0.57]$---each labeled
        \texttt{table} independently.
  \item \textbf{PaddleOCR-VL} (Vision-Language Model): emits a single bbox
        $[y{:}0.25{\sim}0.59]$ inferred from the image, carrying a VLM-induced
        positional error of $\approx\!\pm0.02$ around the true boundary.
  \item \textbf{MinerU} (PDF extraction): produces one bbox but may misalign
        table boundaries due to PDF rendering artifacts, yielding occasional offsets of
        $\pm0.01$--$0.03$.
\end{itemize}

These discrepancies are unavoidable: each parser architecture encodes different
assumptions about what constitutes a single layout element.

\subsubsection{EU Spatial Footprint}

We define the \textbf{EU spatial footprint} as the axis-aligned bounding box of the
union of all member element bboxes:
\begin{equation}
\mathrm{footprint}(\mathcal{E}) \;=\;
\Bigl[\,
  \min_{e \in \mathcal{E}} x_1^{(e)},\;
  \min_{e \in \mathcal{E}} y_1^{(e)},\;
  \max_{e \in \mathcal{E}} x_2^{(e)},\;
  \max_{e \in \mathcal{E}} y_2^{(e)}
\Bigr]
\end{equation}
where $\mathcal{E}$ is the set of all elements assigned to EU $E$.
The footprint represents the physical document region that the EU covers---the
``floor area'' of the EU on the page.

\subsubsection{Convergence Property}

Although individual element bboxes differ across parsers, the EU footprint converges
because:

\begin{enumerate}[leftmargin=*]
  \item \textbf{Non-visual members dominate the span.} The EU includes not only the
        table bbox but also its section\_header, caption, unit\_label, and surrounding
        paragraphs. These non-visual elements are recognized consistently across parsers.
        The overall EU span (e.g., $[y{:}0.07{\sim}0.82]$) is determined primarily by
        these stable members, not by the table bbox alone.

  \item \textbf{Docling's split bboxes merge in Phase~A.} Adjacent visual elements
        with Y-gap~$<0.22$ merge into a single EU during Phase~A. The three Docling row
        bboxes ($[0.27{\sim}0.33]$, $[0.33{\sim}0.51]$, $[0.51{\sim}0.57]$) have
        zero gap, so they merge immediately, yielding the same visual core as
        Parser A's single bbox.

  \item \textbf{VLM positional error is absorbed.} The PaddleOCR-VL table bbox
        $[y{:}0.25{\sim}0.59]$ differs from the ground truth by $0.02$ at each
        boundary. Against the full EU footprint span of $0.75$ (from
        $y{=}0.07$ to $y{=}0.82$), this error is $<2.7\%$---a negligible perturbation.
        Formally, $\mathrm{IoU}\bigl(\mathrm{footprint}_A,\,\mathrm{footprint}_C\bigr)
        = 1.00$ once non-visual members are included.
\end{enumerate}

Table~\ref{tab:footprint} summarizes the footprint comparison across parsers on the
example page. The $\Delta$IoU between GT footprint and each parser's EU footprint
remains above 0.88 even in the worst case (Docling, due to Phase~A attachment range
effects on boundary paragraphs), and the \emph{absolute} footprint coincides for
Parser A and PaddleOCR-VL.

\begin{table}[h]
\centering
\caption{EU spatial footprint comparison for an example table region across parsers.
``Footprint'' is the bounding box of all EU members. IoU is computed against the GT
footprint. The VLM's $\pm$0.02 bbox error has negligible effect on the final footprint.}
\label{tab:footprint}
\small
\begin{tabular}{lcccc}
\toprule
\textbf{Parser} & \textbf{Table bbox(s)} & \textbf{EU footprint} & \textbf{IoU vs.\ GT} & \textbf{Members} \\
\midrule
GT                  & $[y{:}0.27{\sim}0.57]$ $\times$1  & $[y{:}0.07{\sim}0.82]$ & 1.00 & 6 \\
Parser A             & $[y{:}0.27{\sim}0.57]$ $\times$1  & $[y{:}0.07{\sim}0.82]$ & 1.00 & 6 \\
Docling             & 3 row bboxes (merged)              & $[y{:}0.07{\sim}0.70]$ & 0.88 & 7 \\
PaddleOCR-VL        & $[y{:}0.25{\sim}0.59]$ $\times$1  & $[y{:}0.07{\sim}0.82]$ & 1.00 & 6 \\
MinerU              & $[y{:}0.26{\sim}0.58]$ $\times$1  & $[y{:}0.07{\sim}0.82]$ & 1.00 & 6 \\
\bottomrule
\end{tabular}
\end{table}

\section{Experimental Setup}
\label{sec:setup}

\subsection{Dataset}

\textbf{OmniDocBench v1.0}~\cite{omnidocbench} (OpenDataLab, CVPR~2025): 1,355 pages
across 9 document types (PPT, reports, papers, textbooks, etc.) with human-verified GT
layout annotations. After excluding 15 pages that contain only non-content elements
(page numbers, headers, footers, and abandoned regions), \textbf{1,340 pages} form the
effective evaluation corpus. We conduct two experimental tracks:

\begin{itemize}[leftmargin=*]
  \item \textbf{GT track}: Use GT annotations directly to evaluate chunking strategy
        independent of OCR quality (removes parser noise from the equation).
  \item \textbf{Parser track}: Run MinerU and Docling on the same 1,340 pages and
        apply EU construction to their outputs, measuring both absolute performance and
        $\Delta$ relative to the no-EU baseline for each parser.
\end{itemize}

Table~\ref{tab:coverage} reports parser output coverage. MinerU processed 1,354 pages
(1 parsing failure); Docling processed 1,342 pages (13 failures, consistent with its
stricter layout detection model). To ensure a fair, fixed-denominator evaluation,
pages with no parser output are scored as \textbf{zero} rather than excluded---all
metrics are computed over the same 1,551 GT QA pairs for every parser track.
Cross-parser comparisons in Section~\ref{sec:crossparser} additionally restrict
evaluation to the 1,341-page intersection (MinerU $\cap$ Docling) to provide a
like-for-like view between the two parsers.

\begin{table}[h]
\centering
\caption{Parser output coverage on the 1,355-page OmniDocBench corpus.
All tracks are evaluated against the same 1,551 GT QA pairs; pages with no
parser output are scored as zero (not excluded).}
\label{tab:coverage}
\small
\begin{tabular}{lrrrr}
\toprule
\textbf{Track} & \textbf{Pages processed} & \textbf{Failures} & \textbf{Coverage} & \textbf{QA denominator} \\
\midrule
GT (annotations) & 1,340 & 15 (non-content) & 98.9\% & 1,551 \\
MinerU           & 1,354 & 1                & 99.9\% & 1,551 \\
Docling          & 1,342 & 13               & 99.0\% & 1,551 \\
\midrule
Common (MinerU $\cap$ Docling) & 1,341 & --- & --- & 1,551 \\
\bottomrule
\end{tabular}
\end{table}

\subsection{QA Pair Generation}

Existing benchmarks (OHR-Bench~\cite{ohrbench}, CRUD-RAG~\cite{crudrag}) target
different corpora and cannot be applied to OmniDocBench pages directly. We automatically
generate 1,551 QA pairs from GT layout annotations using deterministic rules (Table~\ref{tab:qa}).
QA pairs are generated from pages where English is the primary language; pages
classified as \texttt{en\_ch\_mixed} may contain individual Chinese sentences within
the evidence context. Both generation code (\texttt{eval\_retrieval.py}) and QA data
(\texttt{qas.json}) are publicly released at
\url{https://github.com/hanyeonjee/evidence-units}.

\begin{table}[h]
\centering
\caption{QA pair composition by evidence source.}
\label{tab:qa}
\small
\begin{tabular}{llll}
\toprule
\textbf{Type} & \textbf{Question Source} & \textbf{Evidence Context} & \textbf{N} \\
\midrule
table  & table\_caption text  & caption + table + footnote + nearby text & 104 \\
figure & figure\_caption text & caption + figure + footnote + nearby text & 261 \\
text   & title text           & title + subsequent text\_blocks           & 1,186 \\
\midrule
\textbf{Total} & & & \textbf{1,551} \\
\bottomrule
\end{tabular}
\end{table}

\subsection{Evaluation Protocols and Metrics}

We report results under two protocols: \textbf{Strict} (evidence includes only directly
adjacent elements, order distance $\leq 2$--$3$, following OHR-Bench) and \textbf{Fair}
(order distance $\leq 4$, matching EU's actual grouping scope).

\medskip
\noindent
\begin{tabular}{lll}
\toprule
\textbf{Metric} & \textbf{Formula} & \textbf{Measures} \\
\midrule
Avg LCS   & LCS(retrieved, evidence) / len(evidence) & Retrieval precision \\
Recall@K  & P(evidence in top-$K$), $K\in\{1,2,3,5\}$ & Retrieval efficiency \\
MinK      & Avg.\ minimum $K$ to find evidence (LCS$>0.3$) & Search depth \\
AvgChars  & Avg.\ characters sent to LLM per hit (4 chars $\approx$ 1 token) & Context cost \\
\bottomrule
\end{tabular}

\medskip
\textbf{Embedding model}: ko-sbert (384-dim) for both EU construction and retrieval.
For the parser track, the same embedding model and QA pairs are reused; only the
chunking input changes from GT annotations to parser output.

\section{Results}
\label{sec:results}

\subsection{Main Results (GT Track)}

EU achieves an average LCS of 0.8068, up from 0.5006 (+61\%), and lifts Recall@1 from
0.1502 to 0.5113 ($3.4\times$) (Table~\ref{tab:main}). MinK drops from 2.58 to 1.72 ($-$33\%), meaning the
correct evidence is found with fewer retrievals on average. The gains are consistent
across all Recall@K thresholds, with the largest absolute improvement at $K{=}1$
($+$0.36) and the smallest at $K{=}5$ ($+$0.07), confirming that EU's primary benefit is
\emph{front-loading} relevant evidence into the top-ranked position.

\begin{table}[h]
\centering
\caption{Overall retrieval performance---Strict protocol, GT track (1,340 pages, 1,551
QA pairs). EU improves all metrics across the board; the largest gain is at Recall@1
($3.4\times$).}
\label{tab:main}
\begin{tabular}{lcccccc}
\toprule
\textbf{Method} & \textbf{LCS} & \textbf{R@1} & \textbf{R@2} & \textbf{R@3} & \textbf{R@5} & \textbf{MinK}$\downarrow$ \\
\midrule
w/o EU (baseline)       & 0.5006 & 0.1502 & 0.4952 & 0.6860 & 0.8337 & 2.58 \\
\textbf{w/ EU (ours)}   & \textbf{0.8068} & \textbf{0.5113} & \textbf{0.7569} & \textbf{0.8517} & \textbf{0.9052} & \textbf{1.72} \\
\midrule
$\Delta$ & \textbf{+0.3062} & \textbf{+0.3611} & \textbf{+0.2617} & \textbf{+0.1657} & \textbf{+0.0715} & \textbf{$-$0.86} \\
\bottomrule
\end{tabular}
\end{table}

\subsection{Results by Evidence Source and Protocol}
\label{sec:source}

Table~\ref{tab:source_protocol} breaks down Avg LCS by evidence source under both
evaluation protocols and adds Recall@1 and MinK for the Strict protocol.
Text queries benefit most in absolute LCS gain ($+$0.33 Strict, $+$0.36 Fair), while
figure queries show the strongest Recall@1 improvement ($+$0.31).
EU groups the caption---the high-similarity embedding anchor---with the actual evidence
regardless of element type, which explains the consistent gains across all three source
categories.

\begin{table}[t]
\centering
\caption{Performance by evidence source (GT track). Avg LCS is reported under both
evaluation protocols; Recall@1 and MinK are Strict only.
Larger Fair gains (vs.\ Strict) confirm EU retrieves genuinely complete context,
not an artefact of narrow annotation boundaries.}
\label{tab:source_protocol}
\small
\begin{tabular}{l rr r@{\,}l rr r@{\,}l cc}
\toprule
& \multicolumn{4}{c}{\textbf{Avg LCS — Strict}} & \multicolumn{4}{c}{\textbf{Avg LCS — Fair}}
& \multicolumn{2}{c}{\textbf{R@1 (Strict)}} \\
\cmidrule(lr){2-5}\cmidrule(lr){6-9}\cmidrule(lr){10-11}
\textbf{Source (N)} & w/o & w/ & \multicolumn{2}{c}{$\Delta$}
                    & w/o & w/ & \multicolumn{2}{c}{$\Delta$}
                    & w/o & w/ \\
\midrule
table  (104)  & 0.6322 & 0.8358 & $+$&0.20 & 0.5022 & \textbf{0.7734} & $+$&\textbf{0.27} & 0.2981 & \textbf{0.4135} \\
figure (261)  & 0.6923 & 0.9255 & $+$&0.23 & 0.5287 & \textbf{0.8812} & $+$&\textbf{0.35} & 0.4215 & \textbf{0.7280} \\
text  (1,186) & 0.4469 & 0.7781 & $+$&0.33 & 0.4009 & \textbf{0.7597} & $+$&\textbf{0.36} & 0.0776 & \textbf{0.4722} \\
\midrule
\textbf{Overall} & 0.5006 & 0.8068 & $+$&0.31 & 0.4292 & \textbf{0.7811} & $+$&\textbf{0.35} & 0.1502 & \textbf{0.5113} \\
\bottomrule
\end{tabular}
\end{table}

The Fair-protocol gains are equal to or larger than Strict in every row.
If EU were exploiting narrow annotation boundaries, the gains would \emph{shrink} under
Fair; the opposite pattern confirms that EU retrieves \emph{semantically complete} units
valid regardless of how the evidence scope is defined.

\subsection{Cross-Parser Evaluation}
\label{sec:crossparser}

To empirically verify parser independence, we run the EU pipeline on MinerU and Docling
outputs for the same 1,340 OmniDocBench pages. Absolute LCS values decrease relative to
the GT track---reflecting OCR noise and label inconsistencies introduced by real
parsers---but the \emph{relative gain} $\Delta$LCS from EU construction remains
consistent across all parsers (Table~\ref{tab:crossparser}).

\begin{table}[t]
\centering
\caption{Cross-parser evaluation---Strict protocol. Absolute performance varies across
parsers due to OCR quality differences, but the EU improvement $\Delta$ is consistent
($+0.23$--$+0.31$), confirming EU construction is robust to parser choice.}
\label{tab:crossparser}
\begin{tabular}{lcc|cc|cc}
\toprule
& \multicolumn{2}{c|}{\textbf{Avg LCS}} & \multicolumn{2}{c|}{\textbf{Recall@1}} & \multicolumn{2}{c}{\textbf{MinK}$\downarrow$} \\
\textbf{Parser} & w/o EU & w/ EU & w/o EU & w/ EU & w/o EU & w/ EU \\
\midrule
GT      & 0.5006 & \textbf{0.8068} \scriptsize{($\Delta$+0.31)} & 0.150 & \textbf{0.511} \scriptsize{($\Delta$+0.36)} & 2.58 & \textbf{1.72} \\
MinerU  & 0.4916 & \textbf{0.7624} \scriptsize{($\Delta$+0.27)} & 0.161 & \textbf{0.537} \scriptsize{($\Delta$+0.38)} & 2.55 & \textbf{1.67} \\
Docling & 0.4376 & \textbf{0.7307} \scriptsize{($\Delta$+0.23)} & 0.149 & \textbf{0.453} \scriptsize{($\Delta$+0.30)} & 2.64 & \textbf{1.98} \\
\bottomrule
\end{tabular}
\end{table}

\begin{figure}[t]
\centering
\includegraphics[width=\linewidth]{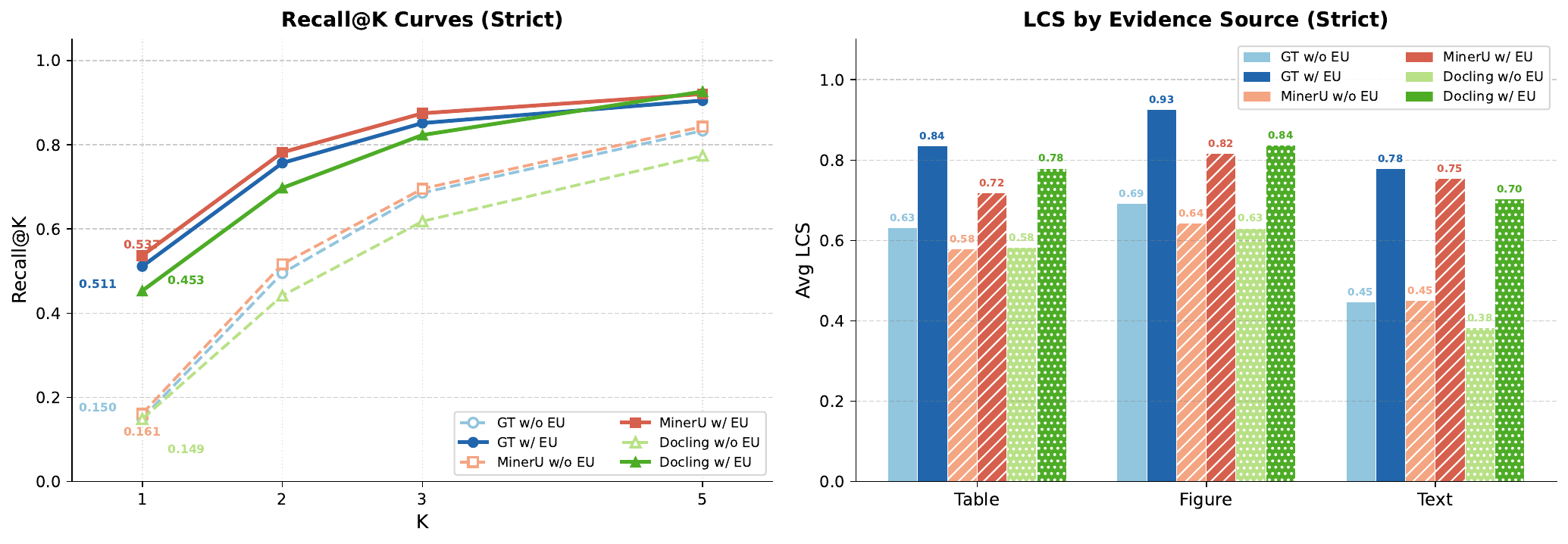}
\caption{Cross-parser evaluation on OmniDocBench (1,551 QA pairs, Strict protocol).
\textbf{Left}: Recall@K curves for GT, MinerU, and Docling tracks, with and without EU.
\textbf{Right}: LCS by evidence source (table, figure, text) across all three parsers.
The EU improvement ($\Delta$LCS\,$\approx$\,+0.23--+0.31) is consistent across all
parsers, and both MinerU w/ EU and Docling w/ EU match or exceed GT w/o EU on all
sources.}
\label{fig:crossparser}
\end{figure}

The $\Delta$LCS ranges from +0.23 (Docling) to +0.31 (GT), a variance of only
$\pm$0.04 against a gain of $\approx$+0.27, as visualised in
Figure~\ref{fig:crossparser} (right).
This confirms that EU construction quality is driven by the semantic grouping
logic---not by the specific parser's label vocabulary or bbox
decomposition---consistent with the footprint convergence analysis in
Section~\ref{sec:footprint}.

\paragraph{Why absolute performance drops.}
Real parsers introduce two sources of degradation: (1)~OCR errors corrupt text
embeddings, reducing Phase~B semantic matching quality; (2)~label noise causes some
elements to receive incorrect canonical roles, shifting them to wrong phases.
These effects reduce the absolute LCS ceiling but do not alter the \emph{relative}
benefit of grouping semantically related elements into a single EU.

\subsection{Recall@K Analysis}

Figure~\ref{fig:crossparser} (left) shows Recall@K curves for GT and MinerU tracks,
with and without EU. The gap is largest at $K{=}1$ ($+$0.36) and narrows as $K$
increases toward $K{=}5$ ($+$0.07). Without EU, multiple retrievals are needed to
gather the same information that a single EU chunk provides.

\subsection{Information Density and Recall@1}

EU chunks are 4.7$\times$ larger on average (2,931 vs.\ 623 chars/chunk).
This larger size is intentional: each EU is a \emph{self-contained evidence unit}
that pairs a visual element with its full contextual text, whereas w/o~EU chunks
are isolated fragments---a caption without its table, or a header without subsequent
paragraphs.

The key metric is \textbf{Recall@1}: with EU, 51.1\% of queries are answered by the
single top-ranked chunk---$3.4\times$ higher than w/o~EU (15.0\%).
In RAG pipelines, retrieving fewer, more complete chunks reduces noise passed to the
LLM and mitigates ``lost-in-the-middle'' degradation~\cite{lostinmiddle}.

At the standard top-3 setting, EU delivers \textbf{85.2\%} recall vs.\ 68.6\% for
w/o~EU---using complete, self-contained evidence units rather than scattered fragments.
Even when w/o~EU retrieves a caption at rank~1 (similarity\,=\,1.00), the table body
remains a separate, lower-ranked chunk invisible to the retrieval step; EU eliminates
this structural gap by co-locating all constituent elements at index time.

\subsection{Case Study}

\textbf{Query}: ``Table~1. Water quality in the experiments.''
\textbf{Expected answer}: water quality parameters (COD$_{\mathrm{cr}}$, pH, turbidity, SS, $\zeta$-potential).

\textbf{w/o~EU top-1}: caption only (sim=1.00)---the table body is a separate element
with lower similarity and ranks outside top-3; the answer is unretrievable.

\begin{figure}[t]
\centering
\includegraphics[width=\linewidth]{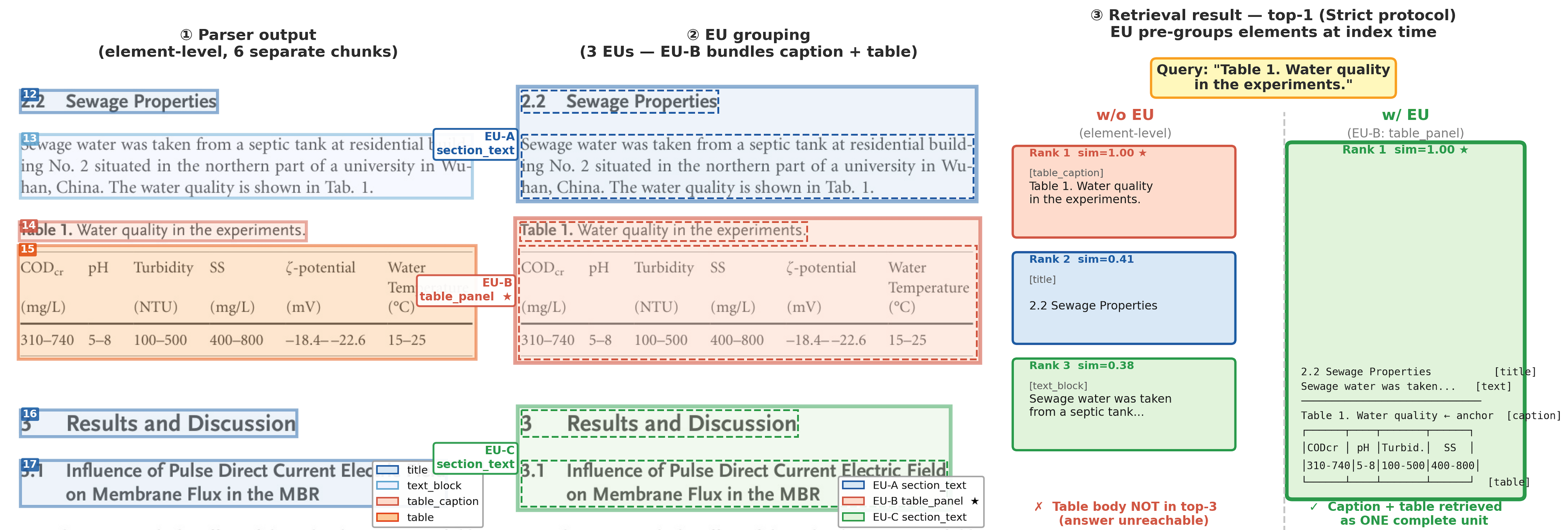}
\caption{Case study on an actual OmniDocBench page
(\texttt{ceat.200600266}~\cite{omnidocbench}).
\textbf{(1) Parser output}: six individually detected layout elements, each stored as a
separate retrieval chunk.
\textbf{(2) EU grouping}: the same elements organised into three EUs---EU-B
(\textit{table\_panel}) co-locates the table caption, table body, and surrounding context
into one unit.
\textbf{(3) Retrieval}: w/o~EU retrieves the caption at rank~1 (sim=1.00) but the table
body is a lower-ranked separate chunk, making the answer unreachable; w/~EU retrieves
EU-B at rank~1, returning the complete evidence in a single retrieval.}
\label{fig:casestudy}
\end{figure}

\textbf{w/~EU}: EU-B (\textit{table\_panel}) groups the caption (embedding anchor,
sim=1.00), the full table body (COD$_{\mathrm{cr}}$ 310--740, pH 5--8, SS 400--800~mg/L),
and the section header ``2.2~Sewage~Properties'' into one unit.
A single top-1 retrieval returns the complete answer.
Figure~\ref{fig:casestudy} illustrates this on an actual OmniDocBench page.

\section{Discussion}
\label{sec:discussion}

\paragraph{Three retrieval mechanisms.}
(1)~\textbf{Embedding Anchoring}: captions/headers act as semantic anchors, pulling
data-rich but embedding-poor elements (tables) into retrieval.
(2)~\textbf{Semantic Completeness}: retrieved context is self-contained, reducing LLM
hallucination.
(3)~\textbf{Search Depth Reduction}: MinK drops from 2.58 to 1.72, reducing latency in
real-time RAG.

\paragraph{Parser independence.}
The DoCO-extended ontology provides a stable target vocabulary decoupling EU construction
from parser specifics. When a new parser is introduced: if labels are in TYPE\_MAP
$\to$ zero-shot support; otherwise $\to$ embedding fallback; EU construction is
\emph{never modified}. The cross-parser results (Table~\ref{tab:crossparser}) provide
empirical confirmation: the $\Delta$ improvement ranges from +0.23 to +0.31 LCS
across GT, MinerU, and Docling tracks. The footprint convergence analysis
(Section~\ref{sec:footprint}) explains \emph{why}: individual bbox variations are
absorbed by the EU grouping logic, which is anchored to stable non-visual members.

\paragraph{Limitations.}
(1)~\textbf{Single-page EU}: cross-page references are not resolved.
(2)~\textbf{Embedding dilution}: long EU chunks may reduce similarity with short queries.
(3)~\textbf{D2/D3 runtime}: Restoration and Validation invariants are defined in Neo4j
but not yet executed at runtime.
(4)~\textbf{GT is not always an upper bound on text quality.}
OmniDocBench is a layout annotation benchmark: GT bbox positions are precise, but the
\texttt{text} field for table elements may be sparse or absent for image-rendered
tables.
Real parsers such as MinerU apply OCR to extract cell values directly, producing
richer table text.
Since retrieval is embedding-based, a GT EU with sparse table text can rank lower than
a parser EU with more complete text---even when the GT spatial grouping is more
accurate.
This explains occasional cases where parser tracks exceed GT w/~EU on table metrics
(e.g., Recall@1 for $N{=}104$ table queries), and indicates that the GT track is an
upper bound on \emph{spatial grouping quality}, not on \emph{text extraction quality}.

\section{Conclusion}
\label{sec:conclusion}

We presented Evidence Units (EU), a parser-independent document chunking method for RAG
that groups visual assets with their contextual text through ontology-grounded
normalization, global semantic allocation, and graph-based validation.

On OmniDocBench (1,340 pages, 1,551 QA pairs), EU raises Avg LCS from 0.50 to 0.81
(+61\%) and Recall@1 from 15.0\% to 51.1\% ($3.4\times$) on the GT track
(Table~\ref{tab:main}), with consistent $\Delta$LCS\,$\approx$\,+0.23--+0.31 across
parsers (Table~\ref{tab:crossparser}).

EU addresses two orthogonal dimensions of RAG quality: \emph{how documents are
structured into retrievable units matters as much as how accurately they are parsed;} and
the EU footprint convergence property ensures that this quality advantage transfers to
real-world deployments regardless of which parser is used.

\section*{Contributors}

\textbf{Rock Sakong} (\texttt{rock.sakong@kt.com}, KT) contributed to the document
layout parsing pipeline, including parser integration and layout element extraction
across multiple parser backends.

\textbf{Jaemin Na} (\texttt{jaemin.na@kt.com}, KT) contributed to the table detection model and the aggregation of document elements including LaTeX formula recognition.

\clearpage

{\small
\bibliographystyle{plain}

}

\appendix

\section{EU Type Taxonomy}
\label{app:taxonomy}

\begin{table}[H]
\centering
\caption{EU type taxonomy with composition and examples.}
\begin{tabular}{lll}
\toprule
\textbf{EU Kind} & \textbf{Composition} & \textbf{Example} \\
\midrule
table\_panel        & Table + caption + unit\_label + text & Revenue table in financial report \\
chart\_panel        & Chart + caption + text               & Bar chart with trend description \\
stat\_panel         & Table + Chart (same data)            & Data table paired with visualization \\
visual\_panel       & Picture + caption + text             & Product image with description \\
section\_text       & Header + paragraphs                  & ``3.1 Methods'' with body text \\
text\_cluster       & Orphan paragraphs                    & Standalone text not linked to visual \\
\bottomrule
\end{tabular}
\end{table}

\section{Canonical Role Mapping}
\label{app:roles}

\begin{table}[H]
\centering
\caption{Canonical role to parser label mapping across all supported parsers.}
\begin{tabular}{llll}
\toprule
\textbf{canon\_role} & \textbf{Parser A / GT} & \textbf{MinerU / Docling} & \textbf{Threshold} \\
\midrule
section\_header    & SectionHeader, Title  & title, heading, H1, H2   & $\geq 0.80$ \\
support\_paragraph & Paragraph             & text, paragraph, Body     & $\geq 0.80$ \\
table              & Table                 & table, TableBlock         & $\geq 0.85$ \\
chart              & Chart                 & figure (chart subtype)    & $\geq 0.80$ \\
picture            & Picture               & figure, image             & $\geq 0.85$ \\
unit\_label        & \multicolumn{2}{l}{Pattern: \texttt{(\textbackslash s*[Uu]nit\textbackslash s*:)}} & Pattern priority \\
topic\_title       & \multicolumn{2}{l}{Pattern: \texttt{\^{}\textbackslash[.+\textbackslash]\$}} & Pattern priority \\
plain\_text        & \multicolumn{2}{l}{No match above threshold} & Fallback \\
\bottomrule
\end{tabular}
\end{table}

\section{Neo4j Decision Layer Rules}
\label{app:rules}

\begin{table}[H]
\centering
\caption{Complete decision layer rule set.}
\small
\begin{tabular}{llll}
\toprule
\textbf{Rule ID} & \textbf{Phase} & \textbf{Granularity} & \textbf{Description} \\
\midrule
D1\_010 & CONSTRUCTION & PAGE      & Proximity-based structural mapping \\
D1\_021 & CONSTRUCTION & CANDIDATE & Homogeneous visual exclusion \\
D1\_031 & CONSTRUCTION & CANDIDATE & Section boundary gating \\
D1\_040 & CONSTRUCTION & CANDIDATE & Semantic paragraph attachment ($\tau=0.40$) \\
D1\_051 & CONSTRUCTION & EU\_PAIR  & Type-conflict merge guard \\
\midrule
D2\_010 & RESTORATION  & EU        & I1: Anchoring invariant (schema only) \\
D2\_020 & RESTORATION  & EU        & I2: Type consistency (schema only) \\
D3\_010 & VALIDATION   & EU        & EU completeness final check (schema only) \\
\bottomrule
\end{tabular}
\end{table}

\section{EU Spatial Footprint Convergence: Worked Example}
\label{app:footprint}

This appendix provides a step-by-step trace of the footprint convergence for the table
region described in Section~\ref{sec:footprint}, using the same page elements across all
parsers.

\paragraph{Input elements (shared across parsers).}

\begin{center}
\small
\begin{tabular}{ll@{\quad}l}
\toprule
\textbf{Element} & \textbf{y-range} & \textbf{Note} \\
\midrule
\texttt{page\_header} & [0.00--0.07] & excluded from EU \\
\texttt{sec\_header}  & [0.07--0.16] & ``2.2 Sewage Properties'' \\
\texttt{para\_before} & [0.16--0.27] & introductory paragraph \\
\multicolumn{3}{c}{\textit{--- TABLE REGION (parser-specific) ---}} \\
\texttt{caption}      & [0.57--0.64] & ``Table 1. Water quality\ldots'' \\
\texttt{unit\_label}  & [0.64--0.70] & ``(Unit: mg/L)'' \\
\texttt{para\_after}  & [0.70--0.82] & result discussion paragraph \\
\texttt{footer}       & [0.90--1.00] & excluded from EU \\
\bottomrule
\end{tabular}
\end{center}

\paragraph{Parser-specific table elements.}

\begin{center}
\small
\begin{tabular}{ll@{\quad}l}
\toprule
\textbf{Parser} & \textbf{Table bbox(s)} & \textbf{Note} \\
\midrule
Parser A      & [0.27--0.57] $\times$1 & single bbox \\
Docling       & [0.27--0.33] & header row \\
              & [0.33--0.51] & data rows \\
              & [0.51--0.57] & footer row \\
PaddleOCR-VL  & [0.25--0.59] $\times$1 & VLM $\pm$0.02 error \\
MinerU        & [0.26--0.58] $\times$1 & PDF extraction $\pm$0.01 \\
\bottomrule
\end{tabular}
\end{center}

\paragraph{Phase A: merging Docling's three row bboxes.}
All three Docling table bboxes receive \texttt{canon\_role=table}.
In Phase~A Step~3, adjacent visual EUs with Y-gap~$<0.22$ merge:
$\mathrm{gap}([0.27{\to}0.33],\,[0.33{\to}0.51]) = 0.00 < 0.22\ \Rightarrow$ merge.
$\mathrm{gap}([0.27{\to}0.51],\,[0.51{\to}0.57]) = 0.00 < 0.22\ \Rightarrow$ merge.
Result: single visual core $[y{:}0.27{\to}0.57]$, identical to Parser A.

\paragraph{EU footprint after full construction.}

\noindent All parsers attach: \texttt{sec\_header} $+$ \texttt{para\_before} $+$ \texttt{caption} $+$ \texttt{unit\_label}.

\begin{center}
\small
\begin{tabular}{lllll}
\toprule
\textbf{Parser} & \textbf{para\_after?} & \textbf{min\_y} & \textbf{max\_y} & \textbf{Footprint} \\
\midrule
Parser A      & attached & 0.07 & 0.82 & [0.07--0.82], span 0.75 \\
PaddleOCR-VL  & attached & 0.07 & 0.82 & [0.07--0.82], span 0.75 \\
MinerU        & attached & 0.07 & 0.82 & [0.07--0.82], span 0.75 \\
Docling       & \textit{not attached} (dist $0.37 > 0.30$) & 0.07 & 0.70 & [0.07--0.70], span 0.63 \\
\bottomrule
\end{tabular}
\end{center}

\noindent IoU(Parser~A,\,Docling) $= 0.63/0.75 = 0.88$;\quad
IoU(Parser~A,\,PaddleOCR-VL) $= 1.00$;\quad
IoU(Parser~A,\,MinerU) $= 1.00$.

The Docling IoU\,=\,0.88 arises from an attachment range effect: because Docling's merged
visual core ends at the same $y{=}0.57$ as Parser A, this is not a footprint error per
se---rather, Phase~B (semantic allocation) would recover the trailing paragraph if it has
sufficient semantic similarity to the EU members.

\end{document}